\documentclass[9pt,twocolumn,twoside]{osajnl}
\journal{ol}
\setboolean{shortarticle}{true}

\title{Spectral multiplexing based on multi-distance lensless imaging}

\author[1]{Qijun You}
\author[1]{Lingshuo Meng}
\author[1]{Yun Gao}
\author[2]{Qing Liao}
\author[1,*]{Wei Cao}
\author[1,3]{Peixiang Lu}

\affil[1]{School of Physics and Wuhan National Laboratory for Optoelectronics, Huazhong University of Science and Technology, Wuhan 430074, China}
\affil[2]{Hubei Key Laboratory of Optical Information and Pattern Recognition, Wuhan Institute of Technology, Wuhan 430205, China}
\affil[3]{Optics Valley Laboratory, Wuhan 430074, China}
\affil[*]{Corresponding author: weicao@hust.edu.cn}

\begin{abstract}
	
 We have demonstrated the capability of spectral multiplexing in multi-distance diffractive imaging, enabling the reconstruction of samples with diverse spectral responses. While previous methods like ptychography utilize redundancy in radial diffraction data to achieve information multiplexing, they typically require capturing a substantial amount of diffraction data. In contrast, our approach effectively harnesses the redundancy information in axial diffraction data. This significantly reduces the amount of diffraction data required and relaxes the stringent requirements on optical path stability.
	  
\end{abstract}

\setboolean{displaycopyright}{true}

\begin{document}
	
	\maketitle
%	\section{Introduction}
   Lensless imaging, as an exquisite solution for high-resolution microscopy, breaks through the limitations of traditional imaging techniques \cite{1}. By abandoning imaging elements such as lenses, the technique successfully gets rid of the constraints of the numerical aperture of the lens and the ribbon sheet on the imaging resolution. Theoretically, its resolution is limited only by the wavelength of incident light, providing a potential possibility for realizing ultra-high-resolution imaging. On the other hand, in the extreme ultraviolet (XUV) and X-ray spectral range, many materials have high absorption properties in these bands. It is extremely difficult to develop corresponding high-quality imaging lenses, which makes lensless diffraction imaging a feasible and efficient solution in the field of extreme ultraviolet imaging \cite{lens2,lens3}.

   The core problem of the lensless imaging technique is how to accurately retrieve the phase information from the recorded diffraction pattern \cite{ph2,ph4}. For the missing phase information, an iterative algorithm can be applied for numerical reconstruction. The phase retrieval algorithm originally originated from the GS algorithm \cite{GS}. Still, to ensure the accuracy and efficiency of phase retrieval, the algorithm often needs to introduce some additional assumptions or constraints in its application. To solve this problem, an iterative approach using more images with different diffraction distances was adopted, resulting in the birth of the multi-distance phase retrieval algorithm \cite{Md, Md2}. This algorithm effectively improves the convergence speed and stability of the algorithm by combining the information of images with multiple diffraction distances, which makes the phase retrieval more accurate and efficient. Another well-known and powerful iterative algorithm, ptychography  \cite{ptychography1,ptychography2}, utilizes a finite illumination probe to scan the sample radially, and the high level of data redundancy due to the overlapping of the irradiated regions contributes to the fast convergence of the iterations. However, conventional lensless imaging techniques usually assume full spatiotemporal coherence of the illumination source. Even relatively small deviations from full coherence can prevent the iterative algorithms from converging correctly, and thus partial coherence of the light source has always been regarded as a nuisance for diffractive imaging.

  Recent studies have proposed new strategies aimed at solving the problem of diffraction pattern ambiguity caused by the illumination of incoherent light sources \cite{broadband1,broadband2,broadband3,broadband4,broadband5,broadband6,broadband7,broadband8,broadband9}. The PIM (ptychographical information multiplexing) based on the ptychography not only overcomes the challenges posed by partial coherence but also reveals more information about the object \cite{broadband3}. The method can reconstruct multiple images simultaneously, thus recovering the spectral response of the object, and is therefore widely used in several fields. However, it is worth noting that the ptychography-based imaging scheme requires the acquisition of a large number of diffraction patterns, which leads to long data acquisition time and places high demands on light source stability. We therefore wish to seek a spectrally multiplexed lensless imaging scheme with fewer frames. In the GS algorithm, recording two diffraction patterns at different distances and iterating back and forth between the two diffraction patterns allows for the reconstruction of lost phase information in the diffraction patterns. It is easy to associate that increasing the number of data acquisition frames between the first and last data acquisition planes is equivalent to providing additional redundant information, which potentially allows us to reconstruct incoherent modes in a partially coherent light field.

  Here we introduce an improved multi-distance phase retrieval algorithm called SMMI (spectral multiplexing multi-distance imaging) for partially coherent sources. By acquiring diffraction patterns at different propagation distances, our method can simultaneously reconstruct multiple exit waves with different frequency modes. Our method enables us to achieve multi-spectral imaging using a reduced amount of diffraction data as compared to the conventional ptycographic approach. It significantly reduces the time needed for data acquisition and relaxes the requirements on optical path stability, thereby enhancing the overall performance and efficiency of the imaging system.
  
	 \begin{figure}[ht]
		\centering\includegraphics[width=1\linewidth]{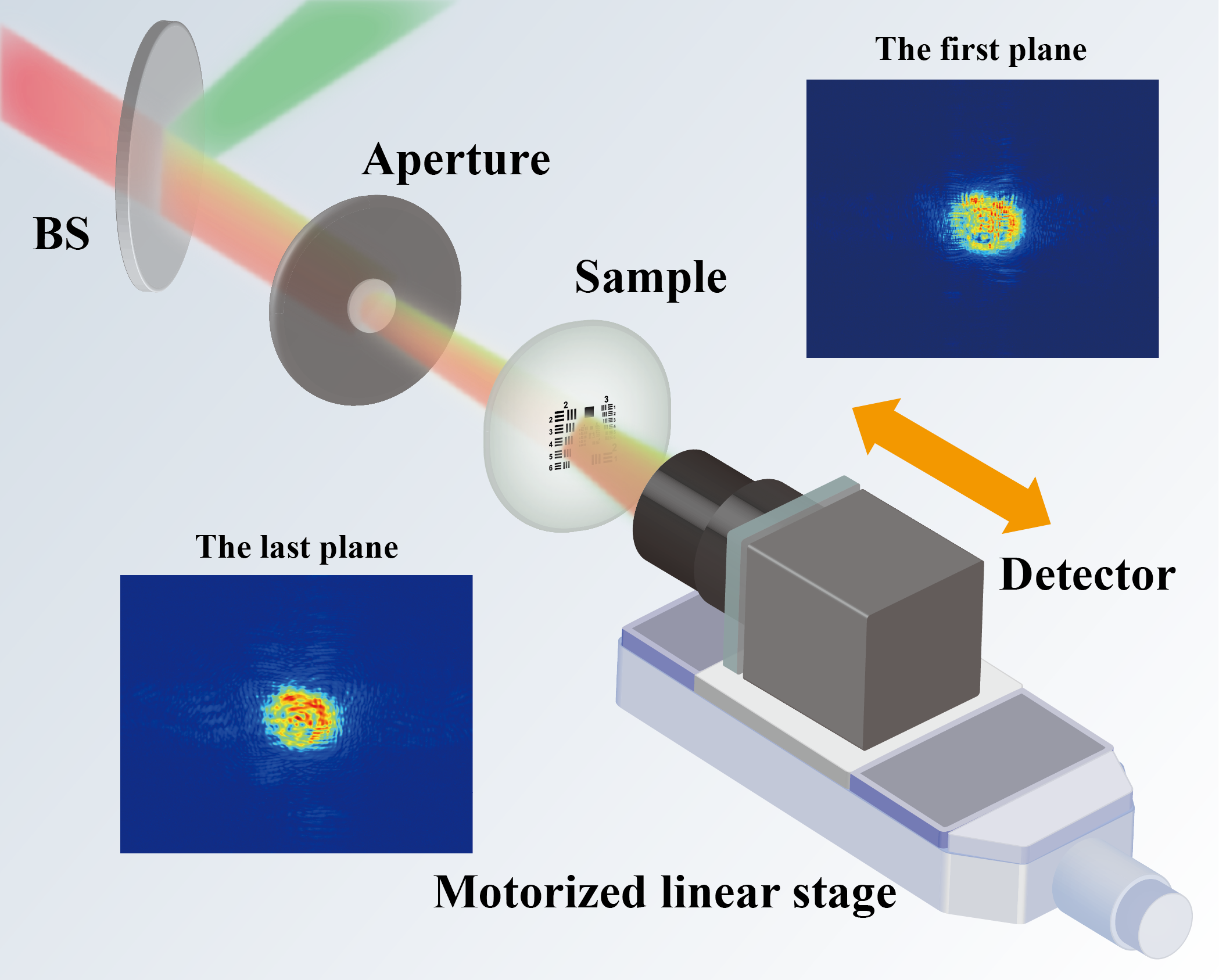}
		\caption{\label{fig1} \footnotesize Schematic of the multi-distance lensless imaging setup. An object is illuminated by a collimated beam to produce a diffraction field. The detector is placed on a motorized stage and moved axially to obtain a series of diffraction patterns at different distances. }
	\end{figure}

     \maketitle
  %   \section{RESULTS AND DISCUSSION}
   To verify the effectiveness of our scheme, we conducted experiments in the visible region, and the experimental setup is shown in Fig.~\ref{fig1}. The red laser ($ \lambda=632 $ nm) and the green laser ($ \lambda=532 $ nm) are respectively expanded and subsequently combined into one beam under the guidance of a beam splitter and projected onto the aperture plane. We use a variable diaphragm as the aperture to flexibly adjust the size of the illumination source. The sample is placed a few centimeters downstream the aperture, and the diffracted light field of the exit wave is recorded by a detector installed on a motorized linear stage (L505, Physik Instrumente, Germany). The detector bit used in the experiments was a CMOS camera (Dhyana, TUCSEN, China) with a pixel size of 6.5 $ \mu $m. The displacement stage was moved axially, 1 mm at a time, for a total of 51 frames of diffraction data. When the illumination laser consists of only two frequency modes, The exit waves of the two lasers can be well reconstructed simultaneously even when less than 10 frames of data are acquired (see details in the Supplementary Material).
   
 Since the incident beam is not exactly perpendicular to the detector plane, this results in a lateral offset of the diffraction data acquired at different locations. Even a small offset, such as a dozen or so pixels, may lead to convergence failure of the phase retrieval algorithm \cite{PZ}. Therefore, it is necessary to preprocess the acquired data for image alignment before feeding them into the phase retrieval algorithm. By calculating the centers of the autocorrelation peaks of the diffracted data at different distances, we can determine the offsets at each position, thus achieving an accurate alignment of the data. This process ensures the effectiveness and accuracy of the subsequent phase retrieval algorithm. The distance from the sample to the detector can be evaluated by a forced auto-focus algorithm \cite{auto-foucs}. In practice, the incoherent diffraction pattern can be regarded as a spectrally weighted incoherent superposition of the diffraction patterns of different wavelength components because the integration time of the detector is much longer than the coherence time between different wavelengths. Based on Wolf's theory of coherent modes \cite{Wolf1982}, a partially coherent wave can be modeled as a set of orthogonal coherent modes propagating independently in free space. This makes it possible to reconstruct the incoherent modes in a partially coherent field.
 
 When the illumination source is non-coherent, then the contribution of all wavelength components to the diffraction pattern must be considered:
 
  \begin{equation}\label{1}
 	\mit\Psi_{\lambda_i}^{j,k}(R)=P^{\lambda_i}_{z}\left\lbrace \psi_{\lambda_i}^{' j,k}\left(r \right) \right\rbrace 
 \end{equation}	

 \begin{equation}\label{2}
 	I^{j,k}=\sum_{i}\left| \mit\Psi_{\lambda_i}^{j,k}(R) \right|^{2}
 \end{equation}	

 The j and k denote the jth iteration and the kth plane, respectively. In our approach, we do not impose any constraints or limitations on the exit wave, and there is no need to input the spectral distribution of the illumination source into the algorithm as a priori information. 

 The diffraction pattern of different wavelengths is measured intensity in the detector plane and is used for intensity constrain in the algorithm:
 
 \begin{equation}\label{3}
 	\mit\Psi_{\lambda_i}^{' j,k}\left(R \right) = \mit\Psi_{\lambda_i}^{j,k}\left(R \right) \frac{ \sqrt{I_{m}^{k}}}{ \sqrt{I^{j,k}}}       
 \end{equation}

$ I_m $ indicates the diffraction data obtained from different planes of collection. Backpropagating the diffracted wavesd back to the near-field plane can obtain the updated exit waves:
\begin{equation}\label{4}
	\psi_{\lambda_i}^{' j,k}\left(r \right) =P^{\lambda_i}_{-z}\left\lbrace \mit\Psi_{\lambda_i}^{' j,k}\left(R \right) \right\rbrace 
\end{equation}	
  The flow chart of the algorthm is shown in the bottom of Fig.~\ref{fig2}.
  
  We performed experiments using a USAF 1951 standard resolution test target ((R1DS1P, Thorlabs, USA) and a biological section as samples (strobile cells), respectively. In order to demonstrate the spectral multiplexing capability of our method, in the first experiment we used the test target as the sample, and we created very different spatial modes for the red and green lasers on the sample by adjusting the position of the aperture. The reconstruction results of the test target are demonstrated in Fig.~\ref{fig2}, where Fig.~\ref{fig2} (a) and (d) show the amplitude distributions of the exit waves obtained using the conventional multi-distance phase retrieval algorithm when only the red or green laser is used as the illumination source. Since the resolution test target has an almost identical spectral response in the visible region, the reconstructed exit waves indicate that the two lasers have very different spatial modes. We then combine both lasers to serve as a single illumination source, by adapting the spectral multiplexing iterative algorithm, we successfully reconstructed the two exit waves with different frequencies simultaneously, as shown in Fig.~\ref{fig2} (c) and (f), which is highly consistent with the reconstruction results from the monochromatic light source in Fig.~\ref{fig2} (a) and (d). These results have demonstrated that our method preserves good sensitivity to identify different spectral modes. It should be noted that in multi-distance phase retrieval algorithms, the ordering of the diffraction data adopted for imposing amplitude constraints will affect the reconstruction results. In Fig.~\ref{fig2} (b) (e), the diffraction data are adopted sequentially, which leads to slow convergence and worse quality of the imaging results. To avoid this issue, we still collect data at different distances with equal spacing in our experiments, but in the algorithm, the data are randomly selected to perform amplitude substitution, which greatly improves both the converging speed and image quality as shown in Fig.~\ref{fig2}(c) (f).
  
 \begin{figure}[ht]
	\centering\includegraphics[width=1\linewidth]{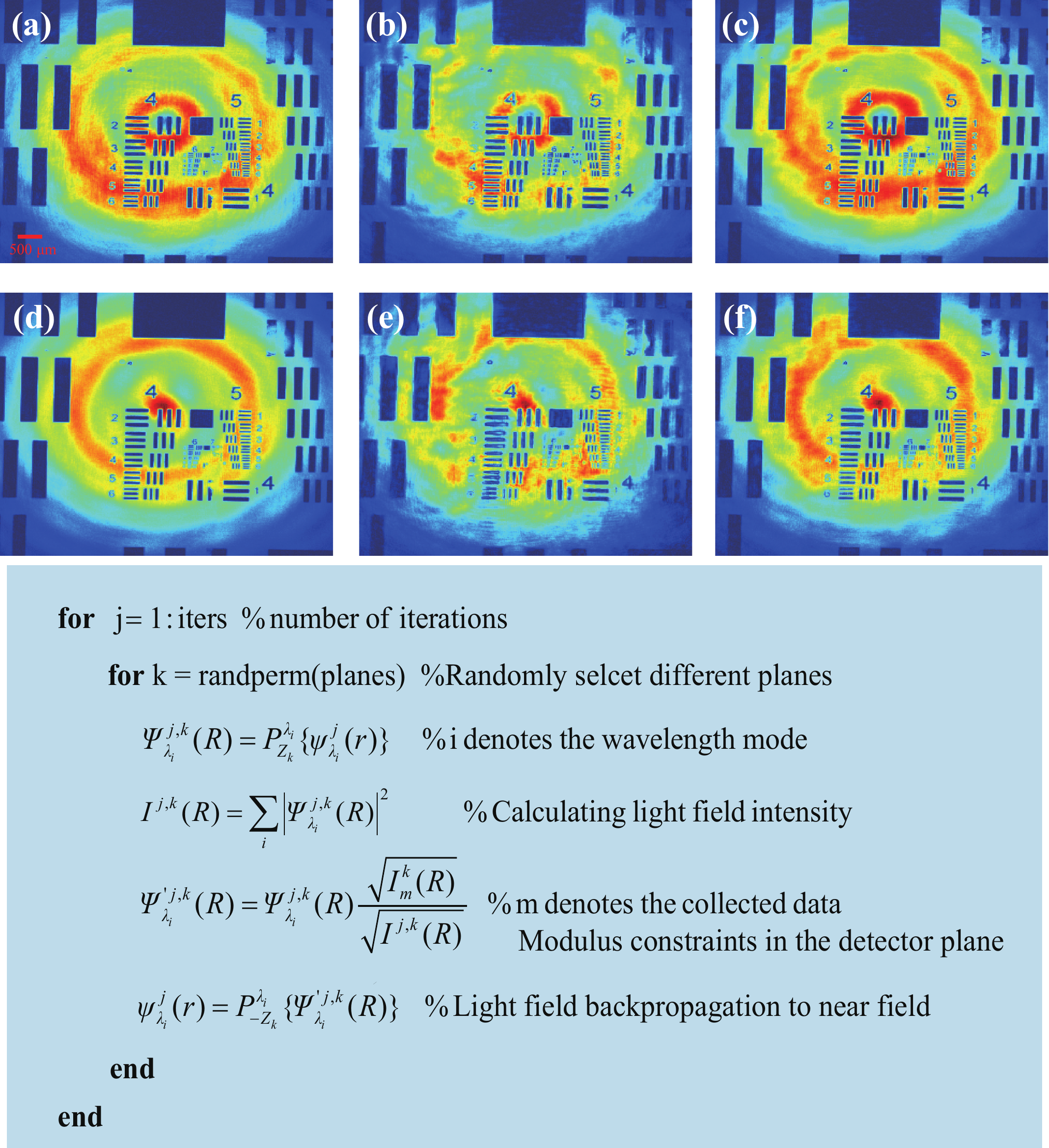}
	\caption{\label{fig2}\footnotesize Experimental results of multi-distance lensless imaging with resolution test targets as the sample. (a) (d) Amplitude distribution of the reconstructed exit wave using red and green light separately. (b) (e) When the combined laser beam is used as an illumination source, the amplitudes of the exit wave of the red and green lasers are reconstructed using the SMMI algorithm, and the amplitude replacements are made in the algorithm by selecting different planes of diffraction data in turn. (c) (f) When the combined laser beam is used as an illumination source, the exit wave amplitudes of the red and green lasers are reconstructed using the SMMI algorithm, which randomly selects different planes of diffraction data for amplitude replacement.}						
 \end{figure}

  The angular spectrum transfer function is used to describe the propagation of the light field in free space. Ideally, the resolution of the reconstructed images is only related to the detector pixel size, which is all 6.5 $ \mu $m. Using the Fresnel diffraction formula, the sampling interval of the reconstruction can be made adaptive to the physical resolution. However special attention needs to be paid to the fact that there is a scaling relationship on the scale of the near-field object function corresponding to different wavelengths. In the SMMI algorithm, the angular spectral function is usually used as a propagator, and after reconstruction of the correct phase of the diffraction data, we can back-propagate using either the angular spectral function or the Fresnel function to obtain the exit wave in the near-field (see Supplementary Material for details).
  
 \begin{figure}[ht]
	\centering\includegraphics[width=1\linewidth]{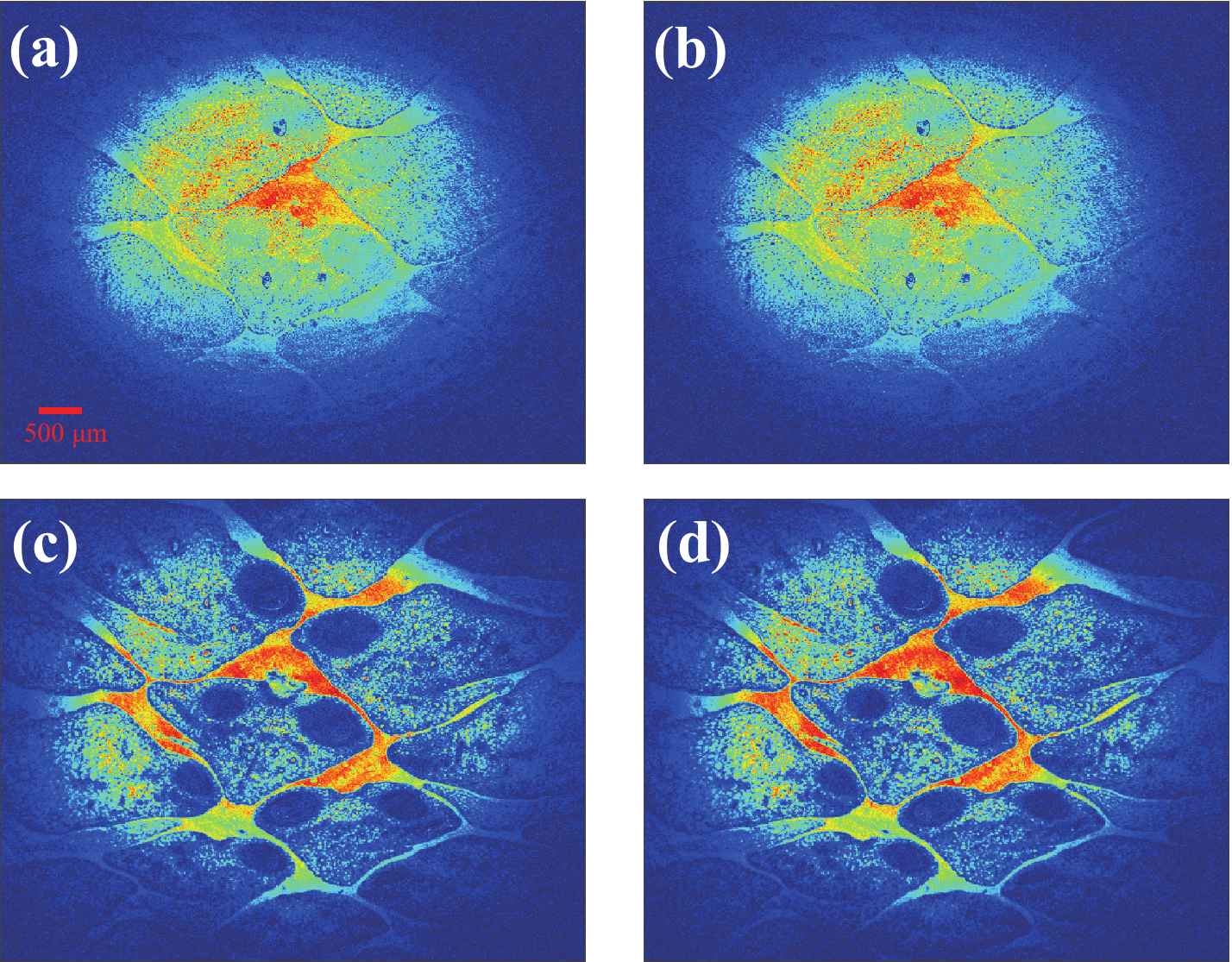}
	\caption{\label{fig3} \footnotesize Results of a multi-distance lensless imaging experiment using a slice of strobile cells as the sample. (a) (c) Amplitude distribution of the reconstructed exit wave using red and green light separately. (b) (d) Illuminated sample after combining red and green light beams. The obtained amplitude distributions of the exit waves are reconstructed simultaneously using our proposed method. }
 \end{figure}

  To demonstrate the versatility of our method, we also carry out diffractive imaging of a biological sample, and the results are presented in Fig.~\ref{fig3}. Among them, Fig.~\ref{fig3} (a) and (c) show the reconstruction results when only the red or green laser is used as the illumination source, respectively. When the sample is illuminated with both lasers presented, the exit wave of the sample can be reconstructed simultaneously using our improved algorithm, and the results are shown in  Fig.~\ref{fig3} (b) and (d). It can be seen that our method can accurately capture the spectral response of the sample. The internal morphology, organizational structure, and composition of macromolecules in biological samples exhibit different optical responses for different wavelengths. These differences provide us with the opportunity to probe deeper into the internal details of biological samples. With our method, biological samples exhibit their unique color gradients when illuminated by a light source composed of multiple wavelengths. These color gradients directly reflect the ability of different components within the sample to absorb light at different wavelengths, resulting in a sharp visual contrast. This contrast not only enhances the visualization of the sample but also greatly simplifies the observation and analysis process.

  The exit wave function is the product of the illumination function and the transmission function of the object, and in many cases, we are more concerned about the actual distribution of the object function, while traditional multi-distance phase retrieval algorithms cannot directly extract the object function. Here, we demonstrate the potential ability of a multi-distance lensless imaging scheme to reconstruct the object function. We first simultaneously reconstructed the illumination function of different wavelengths without the sample in the optical path using the method described previously, as shown in Fig.~\ref{fig4}(a) (d). The sample was then placed in the optical path and diffraction data were collected. The exit wave function of the sample can be written as:
  
\begin{figure}[ht]
	\centering\includegraphics[width=1\linewidth]{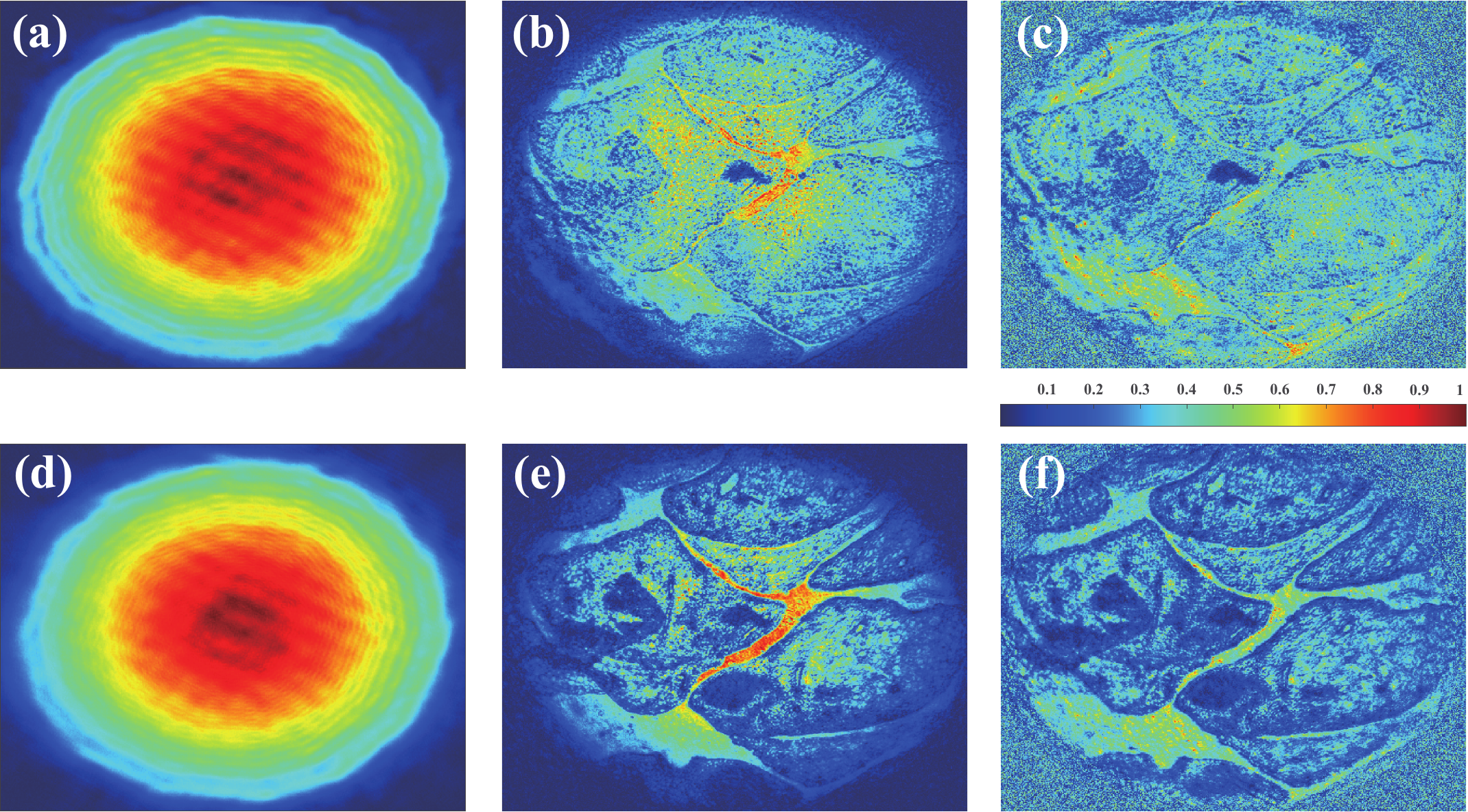}
	\caption{\label{fig4}\footnotesize (a) (d) The sample is moved out of the optical path and the aperture is illuminated by the combined beam of red and green light. The amplitude distribution of the illumination function is simultaneously reconstructed using our proposed method. (b) (e) The sample is placed in the optical path and the amplitude distribution of the exit wave is reconstructed simultaneously. (c) (f) Simultaneous reconstruction of the amplitude distribution of the object function, which exhibits more clearly detailed features compared to the exit wave.}	
\end{figure}

\begin{equation}\label{5}
	\psi_{\lambda_i}^{j,k}\left(r \right) =probe_{\lambda_i}O_{\lambda_i}^{j}(r)
\end{equation}	

The $probe_{\lambda_i}$ is the illumination function obtained by reconstruction. The object function is updated as:

\begin{equation}\label{6}
	O_{\lambda_i}^{j+1}(r)=O_{\lambda_i}^{j}(r)+\alpha\frac{probe_{\lambda_i}^{*}}{\left| probe_{\lambda_i}\right|_{max}^{2} }[\psi_{\lambda_i}^{' j,k}\left(r \right)-\psi_{\lambda_i}^{j,k}\left(r \right)]
\end{equation}
where $\alpha$ is a constant, usually set between 0 and 1. 

  Compared with the reconstructed exit wave (shown in Fig.~\ref{fig4} (b) (e)), the reconstruction result of the object function (shown in Fig.~\ref{fig4} (c) (f)) exhibits clearer and more detailed features. It is worth noting that the phase distribution of the exit wave often does not accurately reflect the true phase information of the sample due to the presence of the illumination function. However, the phase distribution of the object function obtained by our reconstruction can more intuitively demonstrate the phase characteristics of the sample (shown in Fig.~\ref{fig5}).
  
\begin{figure}[ht]
	\centering\includegraphics[width=0.95\linewidth]{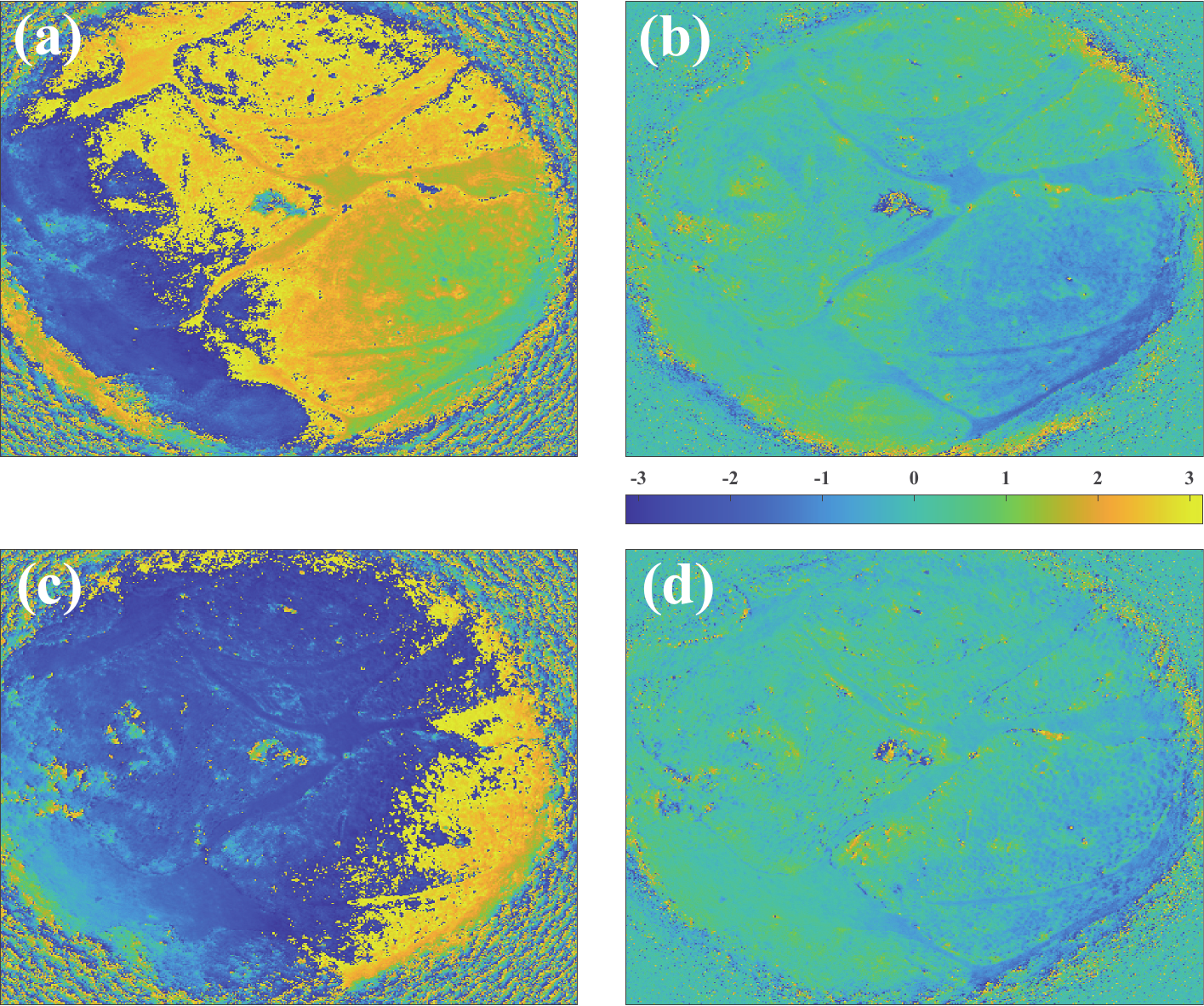}
	\caption{\label{fig5}\footnotesize (a) (c) Retrieved phase distributions from the exit waves shown in Fig.~\ref{fig4} (b) and Fig.~\ref{fig4} (e). (b) (d) Retrieved phase distributions from the exit waves shown in Fig.~\ref{fig4} (c) and Fig.~\ref{fig4} (f).}
\end{figure}

\maketitle
%\section{CONCLUSION}
Overall, we propose a spectral multiplexing multi-distance lensless imaging scheme that is capable of simultaneously reconstructing the exit waves of different modes in a partial coherent light source consisting of multiple frequency components, by applying an improved phase retrieval algorithm without relying on the a priori information of the sample and the illumination source. Compared to PIM, our scheme has reduced data acquisition (a direct comparison between PIM and our scheme can be found in Supplementary Material), faster reconstruction speed, and more relaxed requirements on optical path stability. Although only two illumination modes were used in this paper as proof of principle measurement, the method is in principle compatible with more modes. In addition, we can utilize this method to reconstruct the illumination function at different wavelengths and effectively extract the key information of the object function through the known illumination function. Therefore, it can be predicted that our scheme has a broad application prospect in the fields of bio-imaging and wavefront diagnosis, and is expected to provide new technical means and ideas for research in related fields.

	\begin{backmatter}
		\bmsection{Funding} This research was supported by the National Natural Science Foundation of China under Grant No. 12274158 and No. 12021004.
		
		\bmsection{Disclosures} The authors declare no conflicts of interest.
		
		\bmsection{Data availability} Data underlying the results are not publicly available at this time but may be obtained from the authors upon reasonable request.
		
		\bmsection{Supplemental document} See Supplement 1 for supporting content.
		
	\end{backmatter}
	
	\bibliography{bibliography}
    \bibliographyfullrefs{bibliography}
	
\end{document}